\documentclass{PoS}
\usepackage[latin1]{inputenc}
\usepackage{amsmath}
\usepackage{booktabs}

\newcommand{\npi}{\mbox{$N\pi$} }
\newcommand{\pref}[1]{(\ref{#1})}

\title{Nucleon-pion-state contamination in lattice calculations of the axial form factors of the nucleon}

\ShortTitle{Nucleon-pion-state contributions to the nucleon axial form factors}

\author{\speaker{Oliver B\"ar}\\
       Institut f\"ur Physik, 
       Humboldt Universit\"at zu Berlin\\
       Newtonstra{\ss}e 15,
	D-12489 Berlin,
	Germany \\
       E-mail: \email{obaer@physik.hu-berlin.de}}

\abstract{The nucleon-pion-state contribution to QCD two- and three-point functions used in the calculation of the axial form factors of the nucleon are studied in chiral perturbation theory. For physically small quark masses the nucleon-pion states are expected to dominante the excited-state contamination at large euclidean time separations. To leading order in chiral perturbation theory the results depend on two experimentally well-known low-energy constants only and the nucleon-pion-state contribution can be reliably estimated. The nucleon-pion-state contribution to the axial form factor $G_{\rm A}(Q^2)$ is at the 5 percent level for source-sink separations of 2 fm and shows almost no dependence on the momentum transfer $Q^2$. In contrast, for the induced pseudo scalar form factor $G_{\rm P}(Q^2)$ the nucleon-pion-state contribution shows a rather strong dependence on $Q^2$ and leads to a 10 to 40 percent underestimation of $G_{\rm P}(Q^2)$ at small momentum transfers. Applying the ChPT results to recent lattice data generated by the PACS collaboration we find agreement with experimental data and the predictions of the pion-pole dominance model. 
}

\FullConference{The 36th Annual International Symposium on Lattice Field Theory - LATTICE2018\\
		22-28 July, 2018\\
		Michigan State University, East Lansing, Michigan, USA.}

\begin{document}

\section{Introduction}

Physical point simulations, i.e.\ simulations with quark masses set to their physical values, eliminate the need for a chiral extrapolation, a step that can introduce a significant systematic uncertainty in lattice QCD results. This advantage, however, requires simulations that are numerically still very demanding and need enormous computer ressources. In addition, the notorious signal-to-noise problem  typically gets worse the lighter the pion mass is, thus the euclidean time separations in correlation functions are restricted to rather modest values. At the same time the excited-state contamination due to multi-particle states involving pions grows because the energy gap to the ground state shrinks with lighter pion masses. 

The multi-particle state contamination can in many cases be studied using chiral perturbation theory (ChPT) \cite{Tiburzi:2009zp,Bar:2012ce}. The impact of two-particle nucleon-pion (\npi) states on nucleon observables is phenomenologically interesting, and LO results for the nucleon mass \cite{Bar:2015zwa}, the nucleon axial, scalar and tensor charges \cite{Bar:2016uoj} as well as various first moments of parton distribution functions \cite{Bar:2016jof} can be found in the literature. Recent reviews covering these results are given in \cite{Bar:2017kxh,Bar:2017gqh}.\footnote{In case of the nucleon mass the three-particle $N\pi\pi$-state contribution is also known and negligible in practice \cite{Bar:2018wco}.}  Here preliminary results of an analogous calculation for the \npi contamination in the nucleon axial form factors $G_{\rm A}(Q^2)$ and $G_{\rm P}(Q^2)$ are presented.

\section{The nucleon axial form factors}
The (isovector) nucleon axial form factors are defined by the matrix element of the local isovector axial vector current between single nucleon states,
\begin{equation}\label{DefFF}
\langle N(\vec{p}')|A_{\mu}^a(0)|N(\vec{p})\rangle = \bar{u}(p')\left(\gamma_{\mu}\gamma_5 G_{\rm A}(Q^2) - i \gamma_5\frac{Q_{\mu}}{2M_N}G_{\rm P}(Q^2)\right)\frac{\sigma^a}{2}u(p)\,.
\end{equation}
The nucleon momenta $\vec{p},\vec{p}'$ in the initial and final state imply the euclidean 4-momentum transfer $Q_{\mu}=(i E_{\vec{p}^{\,\prime}} - i E_{\vec{p}},\vec{q}),\, \vec{q} = \vec{p}^{\,\prime}-\vec{p}$. We follow the kinematic setup $\vec{p}^{\,\prime}=0$ that is often chosen in numerical simulations.   $G_{\rm A}(Q^2)$ and $G_{\rm P}(Q^2)$ on the right hand side of \pref{DefFF} refer to the axial and induced pseudo scalar form factors, respectively.

The lattice determination of the two form factors follows a standard procedure.
It is based on the calculation of the nucleon 2-point (pt) function and the nucleon 3-pt function involving the axial vector current, 
where the latter reads \footnote{We follow the conventions chosen in Ref.\ \cite{Capitani:2017qpc}.}
\begin{equation}
C_{3,A^3_{\mu}}(\vec{q},t,t')=\sum_{\vec{x},\vec{y}} \,e^{i\vec{q}\vec{y}}\, \Gamma_{\beta\alpha}\langle N_{\alpha}(\vec{x},t) A_{\mu}^3(\vec{y},t')\overline{N}_{\beta}(0,0)\rangle\,.
\end{equation}
Nucleon interpolating fields $N,\overline{N}$ are placed at time slices $t$ and 0, and the third isospin component of the axial vector current is inserted in between at $t'$. The correlation functions are used to 
form the generalized ratio
\begin{equation}\label{Defratio}
R_{\mu}(\vec{q},t,t') =\frac{C_{3,A^3_{\mu}}(\vec{q},t,t')}{C_2(0,t)}\sqrt{\frac{C_2(\vec{q},t-t')}{C_2(0,t-t')}\frac{C_2(\vec{0},t)}{C_2(\vec{q},t)}\frac{C_2(\vec{0},t')}{C_2(\vec{q},t')}}\,.
\end{equation}
Taking all time separations $t,t',t-t'$ to infinity the ratios approach constants. For the spatial components ($\mu=k=1,2,3$) these are given by
\begin{equation}\label{asymptoticRatios}
R_{{k}}(\vec{q},t,t') \rightarrow \Pi_{{k}}(\vec{q}) = \frac{i}{\sqrt{2E_{\vec{q}}(M_N+ E_{N,\vec{q}})}}\left( (M_N+E_{N,\vec{q}})G_{\rm A}(Q^2) \delta_{3k}-\frac{G_{\rm P}(Q^2)}{2M_N} q_3q_k\right)
\end{equation}
where $E_{N,\vec{q}}$ is the energy of a nucleon with momentum $\vec{q}$. Eq.\ \pref{asymptoticRatios} defines a linear system for  $G_{\rm A}$ and $G_{\rm P}$ which can be easily solved to obtain the two form factors, $\Pi_{{k}}(\vec{q}) \,\rightarrow \, G_{\rm A}(Q^2)\,,\, G_{\rm P}(Q^2)$.

In practice the time separations are finite and far from being asymptotically large. In that case, solving the linear system with $R_{{k}}(\vec{q},t,t') $ instead of $ \Pi_{{k}}(\vec{q})$ we obtain {\rm effective} form factors $G^{\rm eff}_{\rm A}(Q^2,t,t')\,, G^{\rm eff}_{\rm P}(Q^2,,t,t')$. These contain excited-state contributions and depend on both $t$ and $t'$. Quite generally we can write
\begin{equation}
G^{\rm eff}_{\rm A,P}(Q^2,t,t')\, = \,G_{\rm A,P}(Q^2)\bigg[ 1 + \Delta G_{\rm A,P}(Q^2,t,t')\bigg],
\end{equation}
with $\Delta G_{\rm A,P}(Q^2,t,t')$ vanishing for $t,t',t-t'\rightarrow \infty$. The dominant excited-state contribution for large but finite time separations is expected to stem from two-particle \npi states, since these have the smallest energy gap to the single nucleon ground state.

\section{The \npi state contribution in ChPT}

\begin{figure}[t]
\begin{center}
 \includegraphics[scale=0.5]{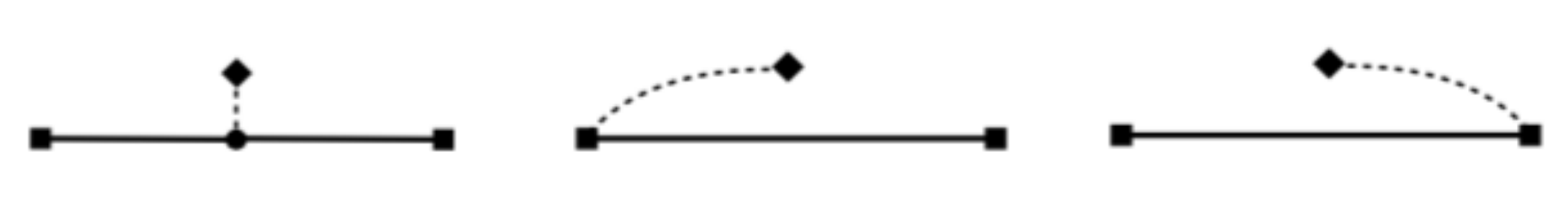} 
\caption{Tree-level Feynman diagrams for the 3-pt function. Solid and dotted lines correspond to nucleon and pion propagators, respectively. Squares, diamonds and circles represent the interpolating nucleon fields, the axial vector current and the interaction vertex (for details see Ref.\ \cite{Bar:2016uoj}).}
\label{fig:treediags}
\end{center}
\end{figure}

The \npi contribution to the effective form factors, $\Delta G^{N\pi}_{\rm A,P}(Q^2,t,t')$, can be computed in ChPT. The calculation is analogous to the one in Ref.\ \cite{Bar:2016uoj} for the  \npi contribution to the axial charge $g_{\rm A} = G_{\rm A}(0)$. In particular, the ChPT setup is independent of the momentum transfer and therefore exactly the same as in Ref.\ \cite{Bar:2016uoj}.
We work to leading order (LO) in SU(2) Baryon ChPT \cite{Gasser:1987rb} and assume isospin symmetry. To this order the chiral effective theory contains the three mass degenerate pions, the mass degenerate proton and neutron fields, and a single interaction vertex which implies the well-known one-pion-exchange-potential between a pair of nucleons.  At LO only two low-energy coefficients (LECs) enter, the axial charge and the pion decay constant. Both are experimentally well determined.

In order to calculate the \npi contribution to the form factors we need to compute the 2-pt and 3-pt functions perturbatively in ChPT. For the 3-pt function this involves the calculation of twelve loop diagrams shown in figure 2 in Ref.\ \cite{Bar:2016uoj}. In addition, the three tree diagrams displayed in fig.\ \ref{fig:treediags} contribute as well. Since their contribution vanishes for zero momentum transfer the tree diagrams were ignored in Ref.\ \cite{Bar:2016uoj}.

All diagrams are computed in the covariant formulation of baryon ChPT. For simplicity, however, the final results are expanded in inverse powers of the nucleon mass. All results shown in the following are based on the leading contribution in this expansion. The calculation of the O($1/M_N$) correction is work in progress. 

With the ChPT results for the correlation functions we form the ratios $R_{k}$ and obtain the effective form factors that contain the \npi state contamination. For a given  source sink separation $t$ we finally determine the plateau estimates 
\begin{eqnarray}
G_{\rm A}^{\rm plat}(Q^2,t)&= \min\limits_{0<t'<t}G_{\rm A}^{\rm eff}(Q^2,t,t')\,,\quad 
G_{\rm P}^{\rm plat}(Q^2,t)&= \max\limits_{0<t'<t} G_{\rm P}^{\rm eff}(Q^2,t,t')\,,
\end{eqnarray}
that are functions of the momentum transfer and $t$. The ChPT calculation is done for a finite spatial volume with spatial extent $L$ in each direction, assuming periodic boundary conditions. This implies discrete momenta $\vec{q}_n$ and 4-momentum transfers $Q^2_n$. 

\begin{figure}[t]
\begin{center}
 \includegraphics[scale=0.60]{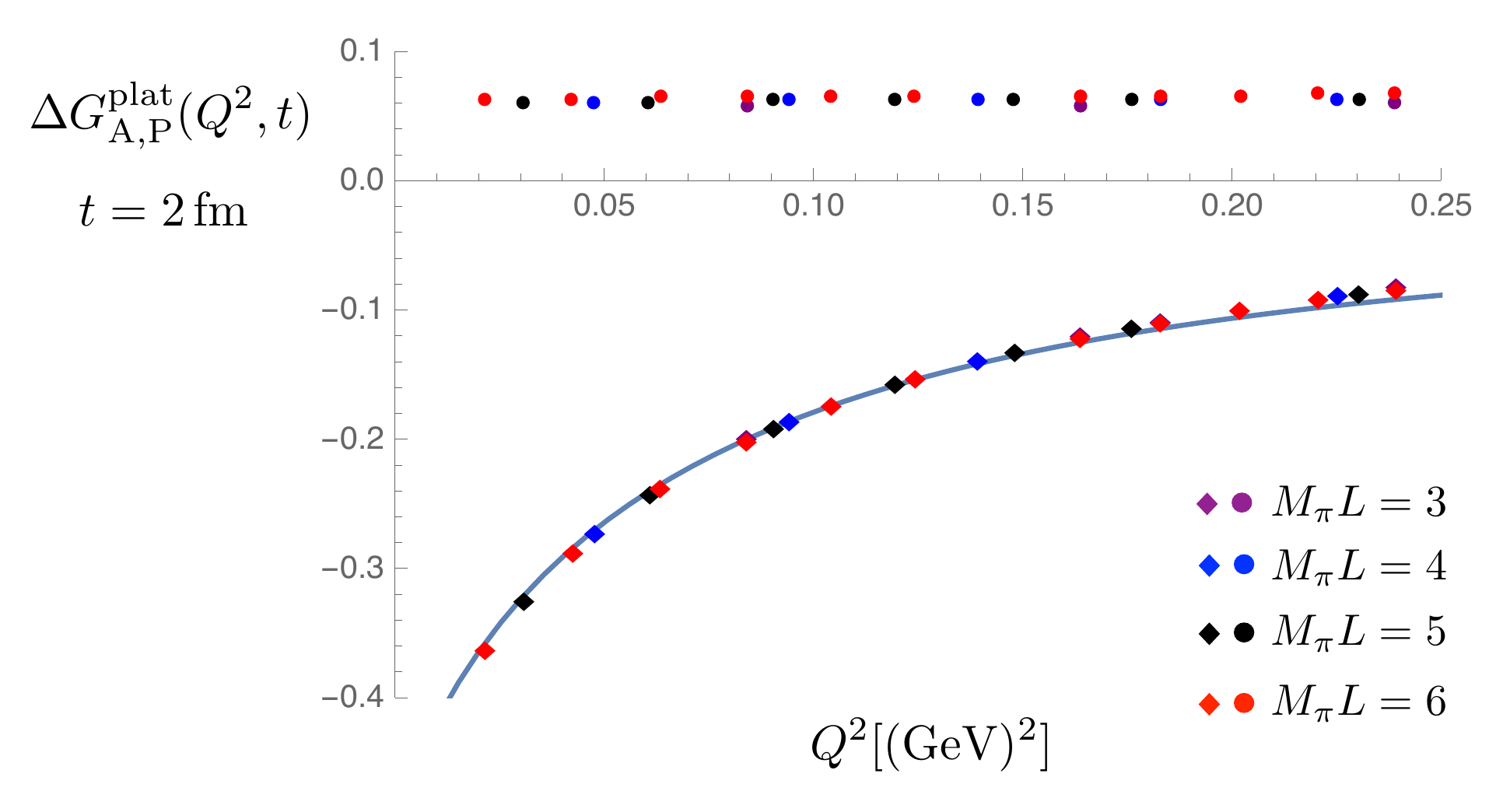}\hspace{1.5cm}\phantom{o}\\
\caption{\label{fig:DeltaG} Results for $ \Delta G^{\rm plat}_{\rm A}$ (dots) and $ \Delta G^{\rm plat}_{\rm P}$ (diamonds) for a source sink separation $t= 2$ fm, and small momentum transfers below $0.25\,\,{\rm (GeV)}^2$. The discrete values for the latter are determined by the size of the spatial volume, and allowed values for four different lattice sizes are shown. The solid line displays the approximate result given in eq.\ \pref{DeltaGPapp}. }
\end{center}
\end{figure}

Figure \ref{fig:DeltaG} shows the relative deviation of the plateau estimates from the true form factor, i.e.\
\begin{equation}
 \Delta G^{\rm plat}_{\rm A,P}(Q^2,t)\equiv \frac{G^{\rm plat}_{\rm A,P}(Q^2,t)}{G_{\rm A,P}(Q^2)} -1\,,
\end{equation}
for a source sink separation of $t=2$ fm and small momentum transfers below $0.25\, {\rm GeV}^2$. Without the $\npi$ contribution $ \Delta G^{\rm plat}_{\rm A,P}$ would be equal to 0. Any deviation from this value is the \npi state contamination in percent. 
Plotted are the results for discrete momentum transfers allowed by various spatial volumes with typical $M_{\pi}L$ values between 3 and 6. In case of the axial form factor (dots) we can read off that the plateau estimates {\em overestimate} $G_{\rm A}$ by about 5\%, essentially independent of $Q^2$. This agrees with the result found for vanishing momentum transfer in \cite{Bar:2016uoj}. In contrast, $G_{\rm P}^{\rm plat}$ {\em underestimates} the pseudo scalar form factor by about 10\% to 40 \% (diamonds), the smaller the momentum transfer the larger the deviation.
Increasing the source sink separation to 3 fm leads to a  smaller \npi contamination of about $+2$\% for the axial form factor, and a 5\% to 20 \% underestimation for the induced pseudo scalar form factor. i.e.\ one roughly gains a factor 1/2.

Note that a final volume effect in these results is small. This is best seen by comparing the results for $M_{\pi}L=3$ and 6, which have some momentum transfers in common. The results for these two volumes have of overlapping symbols. 

Two observations are worth mentioning. Firstly, for some momentum transfers the extraction of the effective form factors from the asymptotic ratios $R_{{k}}(\vec{q},t,t')$
can be done in different ways. For instance, the two momenta $\vec{q}_A = \frac{2\pi}{L}(1,0,1)$ and $\vec{q}_B = \frac{2\pi}{L}(1,1,0)$ imply the same $Q^2$, and the effective form factors can be obtained from three inequivalent linear systems based on three combinations of ratios: i) $R_{3}(\vec{q}_{A},t,t') $ and $R_{3}(\vec{q}_{B},t,t') $, ii) $R_{1}(\vec{q}_{A},t,t') $ and $R_{3}(\vec{q}_{B},t,t') $, iii) $R_{1}(\vec{q}_{A},t,t') $ and $R_{3}(\vec{q}_{A},t,t') $. One might expect that one combination is superior compared to the other two, however, it turns out that all three combinations give practically the same effective form factors.

The second observation concerns the \npi contribution to $\Delta G^{\rm plat}_{\rm P}(Q^2,t)$, which is found to be completely dominated by the left tree diagram in fig.\ \ref{fig:treediags}. The loop diagram contribution turns out to be tiny. As a consequence we can expect ChPT to give more reliable results for $ \Delta G^{\rm plat}_{\rm P}(Q^2,t)$ than for $ \Delta G^{\rm plat}_{\rm A}(Q^2,t)$ since the ``tower'' of narrowly spaced $N\pi$ states does not contribute. In particular, we may expect the ChPT result to be reliable for source sink separations much less than 2 fm.   

For $\Delta G^{\rm plat}_{\rm P}(Q^2,t)$ an excellent approximation for small momentum transfers is given by the simple analytic expression
\begin{equation}\label{DeltaGPapp}
\Delta G^{\rm plat}_P(\vec{q},t)\, \approx\, -\exp\left[-E_{\pi,\vec{q}}\frac{t}{2}\right] \cosh\left[ \frac{q^2}{2 M_N} \frac{t}{2}\right]\,.
\end{equation}
This approximate result is shown in fig.\ \ref{fig:DeltaG} by the solid line. Apparently, the deviation to the diamonds is very small. Note that the approximate result \pref{DeltaGPapp} depends only on the pion and nucleon masses and the source sink separation, not on any other LECs.

\section{Impact on lattice calculations}

In a very recent paper \cite{Ishikawa:2018rew} the PACS collaboration reported lattice data for the two form factors. In contrast to many other collaborations the PACS collaboration simply uses the plateau estimates for the form factors. Thus we can apply the ChPT results presented in the last section to analytically remove the expected $N\pi$ state contamination from the lattice data and check whether better agreement with experimental data and phenomenological models is achieved.

 The PACS results are obtained on a $96^4$ lattice with lattice spacing $a\approx 0.085$ fm and a pion mass $M_{\pi}\approx146$ MeV. The spatial lattice extent $L\approx 8.1$ fm is rather large, corresponding to $M_{\pi}L\approx 6.0$. The source-sink separation equals 15 time slices,  i.e.\ $t\approx1.3$ fm, and the central four time slices with $6\le t/a \le 9$ were used to obtain the plateau estimates. For more simulation details see \cite{Ishikawa:2018rew}.
 
\begin{figure}[t]
\begin{center}
\includegraphics[scale=0.55]{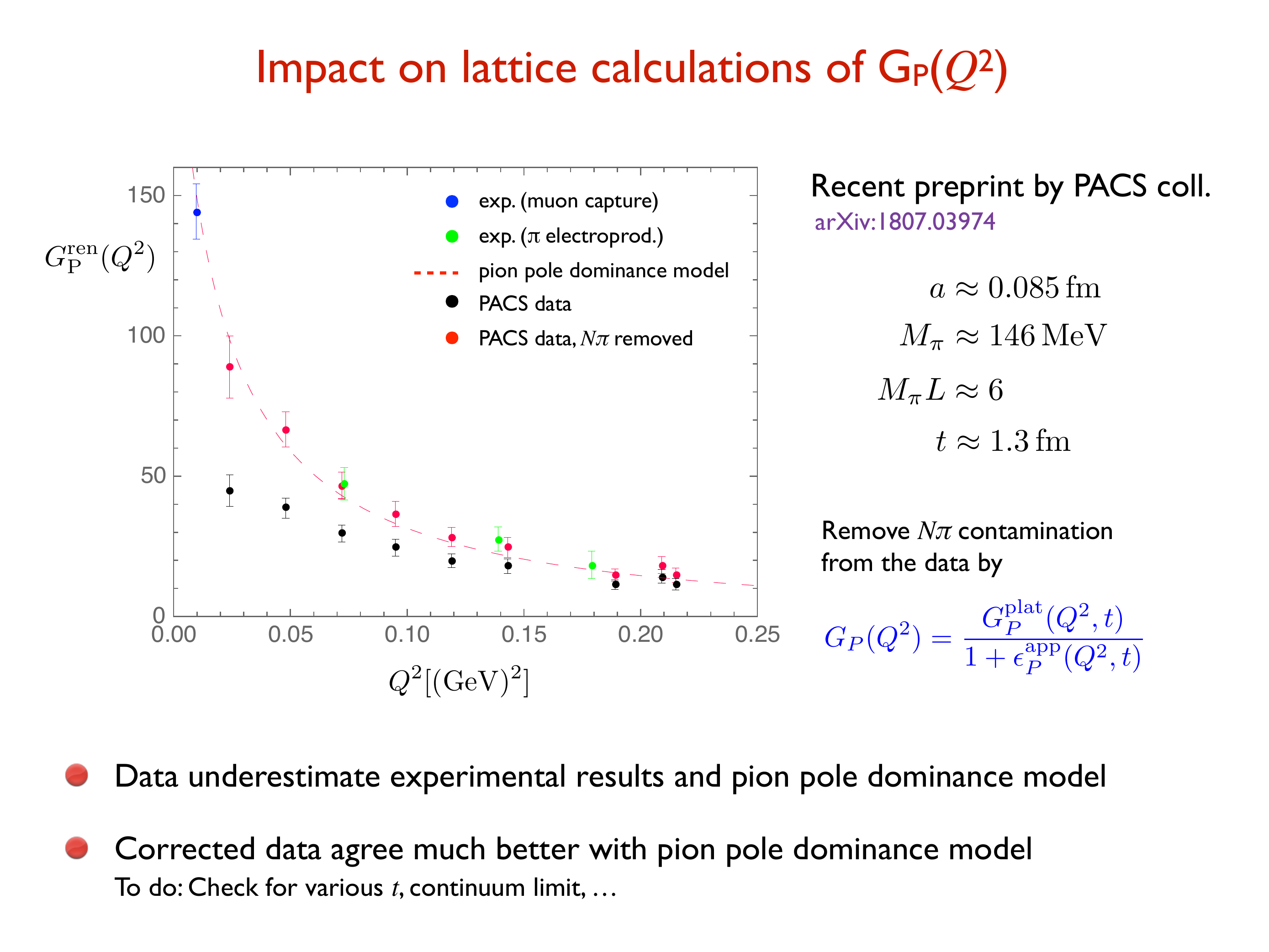}\hspace{1.5cm}\phantom{o}\\
\caption{\label{fig:pacsplot} Lattice data (black data points) reported in Ref.\ \cite{Ishikawa:2018rew} for the momentum transfer dependence of the induced pseudo scalar form factor. The experimental data points stem from muon capture \cite{Andreev:2012fj} and pion-electroproduction \cite{Choi:1993vt} experiments. The dashed line shows the expectation of the ppd model. Removing the \npi state contamination from the lattice data according to eq.\ \pref{pacsdatacorrected} (red data points) leads to much better agreement with the experimental data and the ppd model.  
}
\end{center}
\end{figure}

Figure \ref{fig:pacsplot} shows essentially figure 16 given in \cite{Ishikawa:2018rew}. It displays the numerical PACS results for the renormalized induced pseudo scalar form factor (black symbols) together with existing experimental results (blue and green symbols) and the analytic expectation by the pion-pole-dominance (ppd) model (dashed line). 
In this model the form factors are given by
\begin{equation}
G_{\rm P}(Q^2) \approx \frac{4M_N^2 G_{\rm A}(Q^2)}{Q^2 + M_{\pi}^2}\,,\qquad G_{\rm A}(Q^2) \approxÊ\frac{G_{\rm A}(0)}{(1+Q^2/{M}_{\rm A}^2)^2} \,.
\end{equation}
In Ref. \cite{Ishikawa:2018rew} the value ${M}_{\rm A}^2\approx1.04$ GeV was chosen, stemming from $ r_{\rm A}^2 = 12/ {M}_{\rm A}^2$ with $r_{\rm A}\approx0.67$ fm.  
 
The lattice data are incompatible with the ppd model and the experimental data for small momentum transfers.
As mentioned, the plateau estimates are obtained at a source sink separation $t\approx 1.3$ fm. For such a small time separation we can expect the plateau estimates to differ significantly from the physical values  at $t=\infty$ due to the presence of excited states. However, with the result $\Delta G^{\rm plat}_{\rm P}(Q^2,t)$ given in eq.\ \pref{DeltaGPapp} we can correct the data and analytically remove the anticipated  $N\pi$-state contamination by calculating 
 \begin{equation}\label{pacsdatacorrected}
G^{\rm corr}_{\rm P}(Q^2) \equiv \frac{G^{\rm plat}_{\rm P}(Q^2,t) }{1+ \Delta G^{\rm plat}_{\rm P}(Q^2,t)},
\end{equation}
 setting $t=1.3$ fm. 
 
 The result of this correction is shown in figure \ref{fig:pacsplot} by the red symbols. The corrected lattice data are in much better agreement with the experimental data and the ppd model. In fact, the improvement is better than naively expected. For source sink separations as small as 1.3 fm one would not be too surprised if excited states other than $N\pi$ states also contribute and ``distort" the form factor. The region of applicability of \pref{pacsdatacorrected} needs to be carefully examined. For this data at various source sink separations will be very useful.  At this conference Y.~Kuramashi presented improved PACS results for the form factors that show a clear $t$ dependence \cite{KuramashiLat18}, and it will be very interesting to check whether the lattice data follow the characteristic $t$ dependence given in \pref{DeltaGPapp}. In any case, the main message here is that the \npi state contamination in $G^{\rm plat}_{\rm P}$ causes a softening of the expected ppd behaviour, a feature that has been observed in many lattice results so far.

\section{Summary and outlook}
We presented preliminary results for the \npi excited state contamination in the plateau estimates for the axial form factors of the nucleon. 
At LO in the chiral expansion we find an overestimation for the axial form factor $G_{\rm A}$, essentially independent of $Q^2$.
For the induced pseudo scalar form factor $G_{\rm P}$ ChPT predicts an underestimation that is strongly dependent on the momentum transfer. This systematic effect can qualitatively explain the deviation from the expected pion-pole dominance model behaviour that is typically observed in lattice data for small momentum transfers.

Work on analogous calculations concerning the \npi contamination in the pseudo scalar and electromagnetic nucleon form factors is currently ongoing \cite{OB}.

\section*{Acknowledgements}
Communication with J.~Green, Y.~Kuramashi, K.-F.~Liu, R.~Sommer and T. Yamazaki is gratefully acknowledged. 
This work was supported by the German Research Foundation (DFG), Grant ID BA 3494/2-1.

\end{document}